\pdfoutput=1
\documentclass[12pt,letter]{article}

\usepackage{jheppub}
\allowdisplaybreaks[1]
\usepackage{bm}
\usepackage{subfigure}

\title{Notes on frequencies and timescales in nonequilibrium Green's functions}

\author{Takaaki Ishii}
\affiliation{Department of Physics, University of Colorado, 390 UCB, Boulder, CO 80309, USA}
\affiliation{Center for Theory of Quantum Matter, University of Colorado, Boulder, CO 80309, USA}
\emailAdd{takaaki.ishii@colorado.edu}

\abstract{We discuss the ringdown behavior of the nonequilibrium Green's function in a strongly coupled theory with the holographic dual with a focus on quasinormal-mode equilibration. We study the time resolved spectral function for a probe scalar in Vaidya-AdS spacetime in detail as a complement to the preceding work \cite{Banerjee:2016ray} using further numerical results in very nonadiabatic temperature changes. It is shown that the relaxation of the nonequilibrium spectral function obtained through the Wigner transform is governed by the lowest quasinormal mode frequency. The timescale of the background temperature change is also observed in the frequency analysis. We then consider a toy model motivated by the quasinormal mode behavior and discuss these main features in numerical results are simply realized.}

\begin{document}

\maketitle

\section{Introduction}
\label{sec:intro}

In description of strongly coupled systems by dual gravitational theories, black hole quasinormal modes \cite{Berti:2009kk} play an important role in relaxation. Approach to equilibrium is damping oscillation with the most dominant quasinormal mode frequency, the lowest one in the tower of the modes. The quasinormal mode ringdown has been observed in local operators in time evolution of far-from-equilibrium systems \cite{Chesler:2008hg,Chesler:2009cy,Bantilan:2012vu,Heller:2012km,Buchel:2012gw,Bhaseen:2012gg,Heller:2013oxa,Chesler:2013lia,Fuini:2015hba,Ishii:2015gia,Gursoy:2016tgf,Gursoy:2016ggq}. Calculating nonequilibrium correlation functions has attracted much attention recently \cite{CaronHuot:2011dr,Chesler:2011ds,Chesler:2012zk,Balasubramanian:2012tu,Keranen:2014lna,David:2015xqa,Keranen:2015mqc,Lin:2015acg,Banerjee:2016ray},\footnote{Thermalization of heavy nonlocal operators also has been studied often \cite{Balasubramanian:2010ce,Balasubramanian:2011ur,Aparicio:2011zy,Balasubramanian:2012tu,Buchel:2014gta,Ecker:2015kna}.} and the quasinormal mode behavior has been also discussed.

Recently in \cite{Banerjee:2016ray}, nonequilibrium Green's and spectral functions were computed for a probe scalar in holography, where the Wigner transform was used for the time dependent frequency analysis. Difference in the behaviors in adiabatic and nonadiabatic background temperature changes was mainly discussed. In late time, it would be natural to expect that the ringdown is dominated by the lowest quasinormal mode, but the pattern in the approach to the thermal equilibrium seemed to vary with the Wigner frequency in a complex way in the limited results in that paper. Further examination therefore should be necessary.

In this paper, we study the ringdown of the nonequilibrium spectral function in detail. Firstly in section~\ref{sec:num}, we do numerical computations and show further results in the regime of very nonadiabatic temperature changes. In particular, frequency dependence and timescales in the equilibration are focused on. In appendix~\ref{app:time}, we give comments on the Fourier momentum dependence. In section~\ref{sec:toy}, we introduce a toy model motivated by the quasinormal mode behavior to consider the numerical results from a general perspective. We conclude with a summary and short discussions.

\section{Nonequilibrium spectral function in holography}
\label{sec:num}

In this section, we consider numerical computation of the nonequilibrium spectral function in holography. We continue to use the same model as \cite{Banerjee:2016ray} given by a probe $\Delta=2$ scalar in Vaidya-AdS$_4$ spacetime, but here we provide further results in more nonadiabatic parameters for convenience. Technical details were explained in \cite{Banerjee:2016ray}. Here we begin with a summary and then show results.

We would like to calculate the retarded Green's function for a probe scalar operator in the linear response theory. The linear response relation is given by
\begin{equation}
\langle O(v_2,\bm{x}_2) \rangle_J = - \int dv d^2x \, G_R(v_2,\bm{x}_2;v,\bm{x}) J(v,\bm{x}),
\label{linearresp}
\end{equation}
where $J$ is the source and $G_R$ is the retarded Green's function
\begin{equation}
G_R(v_2,\bm{x}_2;v_1,\bm{x}_1) = -i \theta(v_2-v_1) \langle [O(v_2,\bm{x}_2), O(v_1,\bm{x}_1)] \rangle.
\label{GROO}
\end{equation}
For simplicity, we assume that $\langle O \rangle $ is trivial if $J=0$.

To obtain the Green's function, an impulse response is convenient. Using $J(v,\bm{x})=\delta(v-v_1)\delta(\bm{x}-\bm{x}_1)$, we obtain
\begin{equation}
\langle O(v_2,\bm{x}_2) \rangle_J = - G_R(v_2,\bm{x}_2;v_1,\bm{x}_1).
\end{equation}
Hence, we calculate the response when the delta function source is applied at $(v_1,\bm{x}_1)$. If the system is homogeneous in the $\bm{x}$-directions, the partial Fourier transform gives
\begin{equation}
\langle O(v_2,\bm{k}) \rangle_f = - G_R(v_2,v_1;\bm{k}),
\label{onepointfn_is_GR}
\end{equation}
where the source is $f(v)=\delta(v-v_1)$ and $\bm{k}$ is the momentum conjugate to $\bm{x}_2-\bm{x}_1$.

Once the above Green's function is computed, we turn to time dependent frequency analysis. We simply consider the Wigner transform
\begin{equation}
\widetilde{G}_R(\omega,\bar{t};\bm{k}) = \int dt \, e^{i\omega t}G_R(\bar{t}+t/2,\bar{t}-t/2;\bm{k}),
\label{GRWigner}
\end{equation}
where $t=v_2-v_1$ and $\bar{t}=(v_2+v_1)/2$. The spectral function is given by\footnote{If the background does not change, the Wigner transform reduces to the standard Fourier transform, and the spectral function simply takes the thermal form.}
\begin{equation}
\rho(\omega,\bar{t};\bm{k}) = -2 \, \mathrm{Im} \, \widetilde{G}_R(\omega,\bar{t};\bm{k}).
\end{equation}
In particular, we focus on the approach to the thermal equilibrium,
\begin{equation}
\Delta \rho(\omega,\bar{t};\bm{k}) = \rho(\omega,\bar{t};\bm{k}) - \rho(\omega,\infty;\bm{k}).
\label{deltarho}
\end{equation}

Calculation of the Green's function is done in the gravity dual. Specifically, we adopt the four-dimensional Vaidya-AdS geometry as an ideal nonequilibrium background.\footnote{The Vaidya solution immediately reaches the final temperature as the mass injection, and the geometry itself does not exhibit ringdown by the dynamical quasinormal mode of the metric. See \cite{Bhattacharyya:2009uu,Ebrahim:2010ra} discussing such instantaneous thermalization.} The metric in the ingoing Eddington-Finkelstein coordinates is given by
\begin{equation}
ds^2 = \frac{1}{z^2} \left(- F(z,v) dv^2 - 2 dv dz + d \bm{x}^2 \right), \quad F(z,v)=1-M(v)z^3,
\label{vaidyaads4metric}
\end{equation}
where the AdS radius is set unity. We choose the mass function $M(v)$ as
\begin{equation}
M(v) = M_i + \frac{M_f - M_i}{2}\left(1+\tanh \frac{v}{\Delta v}\right),
\label{massfntanh}
\end{equation}
where $M_i$ and $M_f$ are the initial and final mass density of the black brane, and $\Delta v$ controls the timescale of the mass change. We use units in which the initial temperature is $T_i=3/(4\pi)$, corresponding to $M_i=1$. We set $M_f=8$ so that the final temperature $T_f$ satisfies $T_f/T_i=2$.

We put a probe scalar with the conformal dimension $\Delta=2$ in this nonequilibrium background. We numerically solve the equation of motion and compute the spectral function as in \cite{Banerjee:2016ray}. In particular, the delta function source is numerically introduced as the narrow limit of a normalized Gaussian function. In the following, we discuss the results.

\begin{figure}[t]
\centering
\subfigure[$k=0$: $\omega=2$ (red), 10 (green), 40 (purple)]{\includegraphics[height=4.5cm]{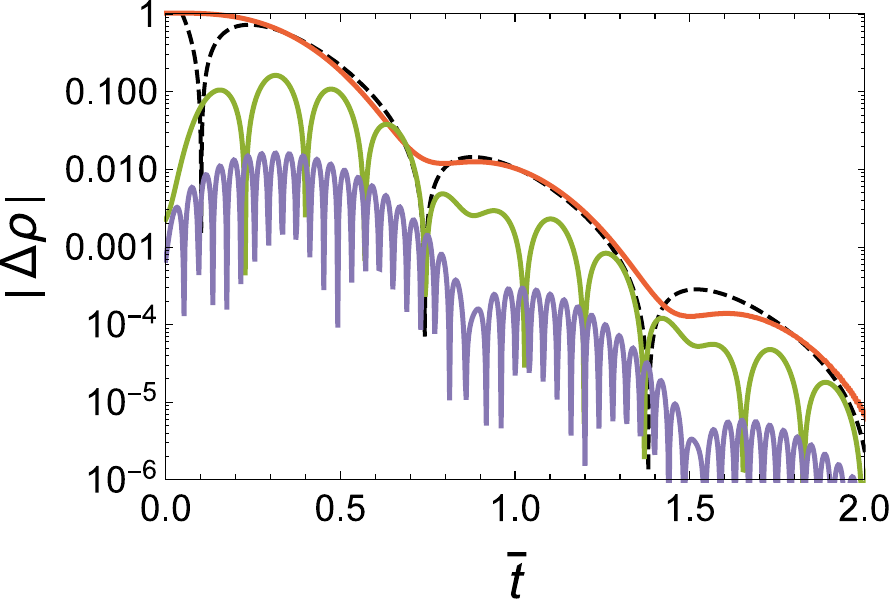}\label{fig:num_k0a002_t}}
\subfigure[$k=0$: $\bar{t}=0.4$ (blue), 0.8 (green), 0.9 (red)]{\includegraphics[height=4.5cm]{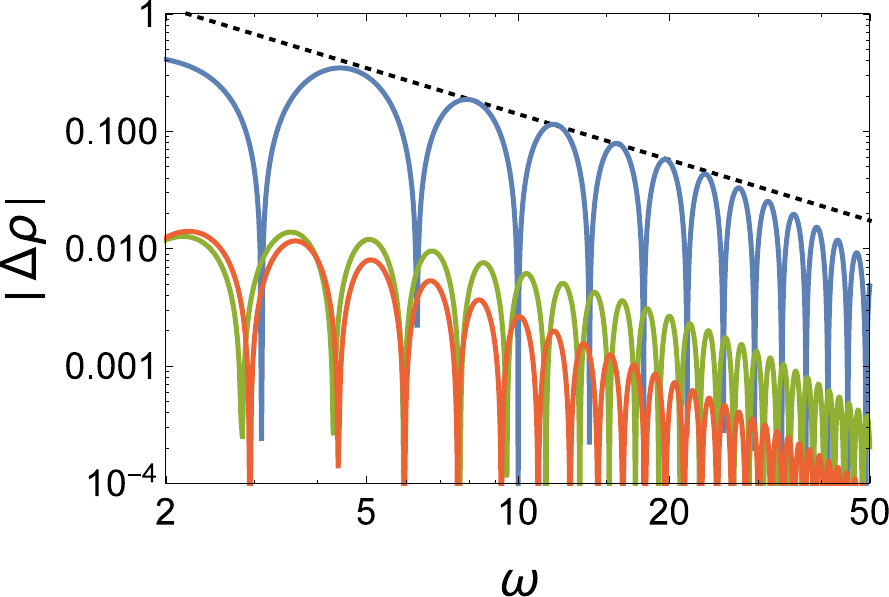}\label{fig:num_k0a002_w}}
\subfigure[$k=4$: $\omega=5$ (red), 12 (green), 36 (purple)]{\includegraphics[height=4.5cm]{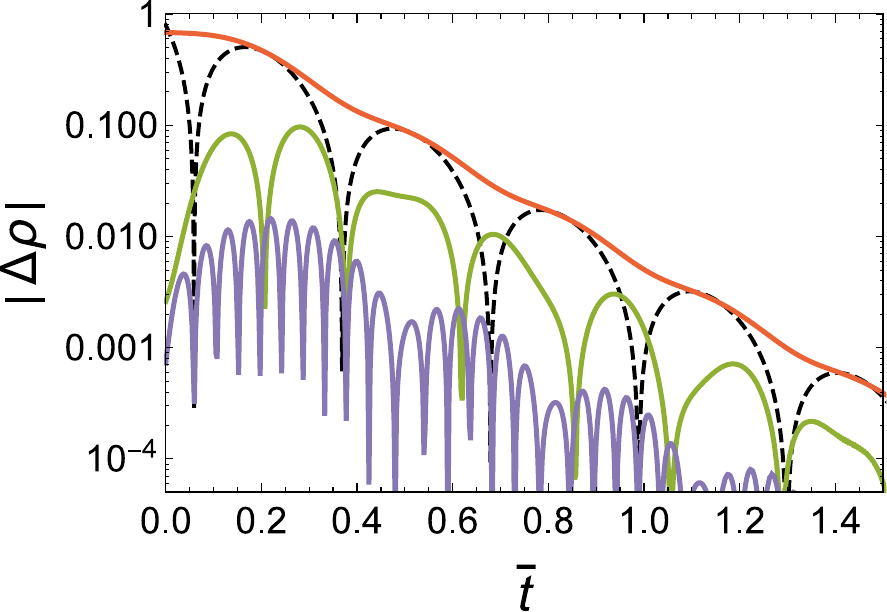}\label{fig:num_k4a002_t}}
\subfigure[$k=4$: $\bar{t}=0.4$ (blue), 1 (green), 1.1 (red)]{\includegraphics[height=4.5cm]{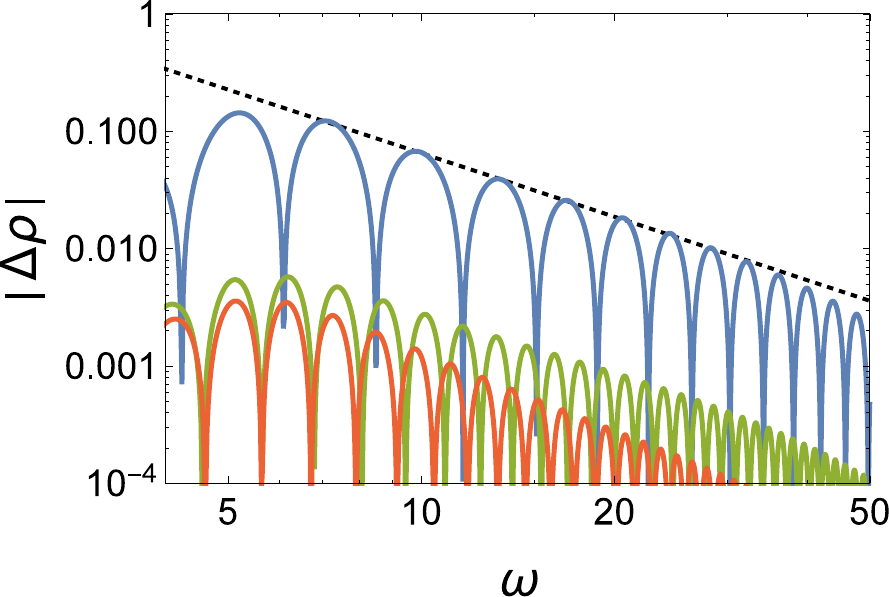}\label{fig:num_k4a002_w}}
\subfigure[$k=8$: $\omega=9$ (red), 20 (green), 45 (purple)]{\includegraphics[height=4.5cm]{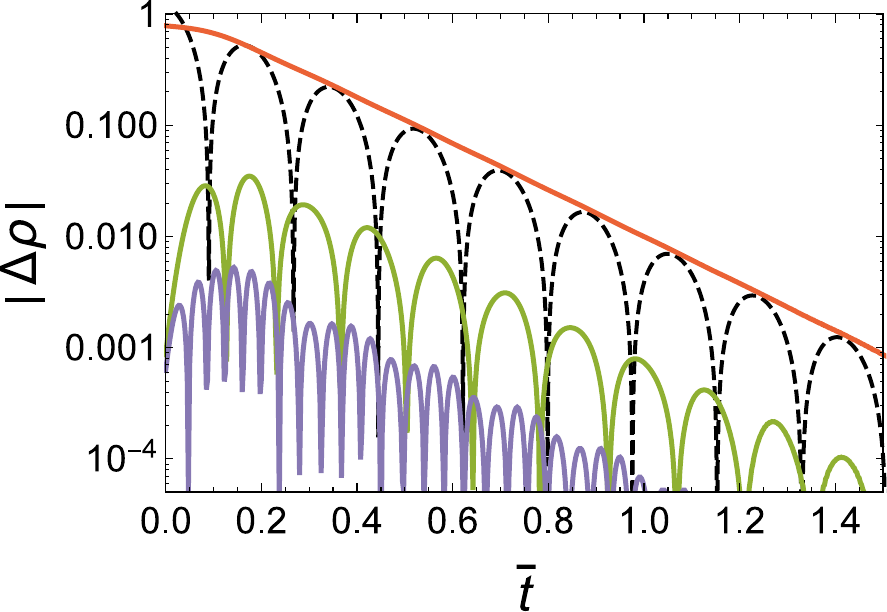}\label{fig:num_k8a002_t}}
\subfigure[$k=8$: $\bar{t}=0.4$ (blue), 0.8 (green), 1 (red)]{\includegraphics[height=4.5cm]{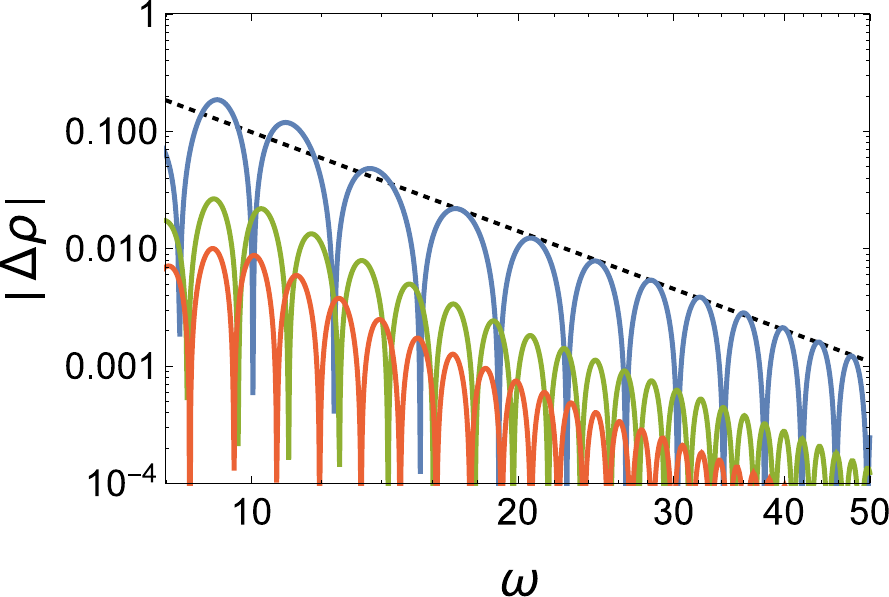}\label{fig:num_k8a002_w}}
\caption{Plots of $\Delta \rho$ when $\Delta v=0.02$. Figures~(a), (c), (e) are for the $\bar{t}$-dependence. The black dashed lines are $5 \sin(2 \omega_f \bar{t} -0.5) e^{-2 \Gamma_f \bar{t}}$, $1.4 \sin(2 \omega_f \bar{t} -0.6) e^{-2 \Gamma_f \bar{t}}$, $1.2 \sin(2 \omega_f \bar{t} -1.6) e^{-2 \Gamma_f \bar{t}}$ in (a), (c), (e), respectively. Figures~(b), (d), (f) are for the $\omega$-dependence. The black dotted lines are $2.8 \, \omega^{-1.3}$, $4.1 \, \omega^{-1.8}$, $62 \, \omega^{-2.8}$ in (b), (d), (f), respectively.}
\label{fig:num_a002}
\end{figure}

\begin{table}[t]
\centering
\begin{tabular}{|c|c|c|}
\hline
$k$ & $\omega_i -i \Gamma_i$ & $\omega_f -i \Gamma_f$\\ \hline
0 & $1.228-1.533i$ & $2.456-3.067i$ \\ \hline
4 & $4.434-1.219i$ & $5.069-2.721i$ \\ \hline
8 & $8.364-1.073i$ & $8.868-2.437i$ \\ \hline
\end{tabular}
\caption{The lowest quasinormal mode frequencies for the initial and final black branes when $T_i=3/(4\pi)$ and $T_f/T_i=2$. The final state momenta are hence $k/(4 \pi T_f/3) = 0,\, 2, \, 4$ from top to bottom.}
\label{QNMfreqtable}
\end{table}

Firstly, we choose $\Delta v=0.02$ at the initial state's momenta $k=0, \, 4$, $8$, where $k \equiv |\bm{k}|$.\footnote{In units of the final state temperature, these momenta correspond to $k/(4\pi T_f/3)=0, \, 2, \, 4$, respectively, since $T_f/T_i=2$.} This $\Delta v$ is smaller than the narrowest one in \cite{Banerjee:2016ray}. Results are shown in figure~\ref{fig:num_a002}. As a reference material, the lowest quasinormal mode frequencies for relevant $k$ are given in table~\ref{QNMfreqtable}, which are computed in the static (Schwarzschild-AdS$_4$) black brane background following the general procedure outlined in e.g.~\cite{Starinets:2002br}.\footnote{Quasinormal modes for (global) AdS black holes were first considered in \cite{Chan:1996yk}. In AdS/CFT, the relation of asymptotically AdS quasinormal modes to the dual field theory was pointed out in \cite{Horowitz:1999jd}.} Figures~\ref{fig:num_k0a002_t}, \ref{fig:num_k4a002_t}, \ref{fig:num_k8a002_t} are for the $\bar{t}$-dependence. For comparison, reference damping oscillations purely by the quasinormal mode frequency are also plotted in black dashed lines. In figures~\ref{fig:num_k0a002_w}, \ref{fig:num_k4a002_w}, \ref{fig:num_k8a002_w}, the $\omega$-dependence at fixed $\bar{t}$ is shown, where each additional black dashed line indicates a power law dependence in the log-log plot.

In the $\bar{t}$-dependence, we see that the equilibration is indeed controlled by the quasinormal mode.
The damping rate is given by the imaginary part of the frequency. Due to the Wigner transform, the time oscillations are subject also to the Wigner frequency $\omega$ as well as the real part of the quasinormal mode frequency. In $\omega \gg \omega_f$ the fastest scale is dominated by $\omega$, but there is the large scale modulation by $\omega_f$.\footnote{In larger $k$, it is necessary to go to $\omega \gg \omega_f$ to clearly see the large scale modulation.} The oscillations are suppressed in intermediate frequencies due to interference of $\omega$ and $\omega_f$. It is around $\omega \simeq \omega_f$ in figures~\ref{fig:num_k4a002_t} and \ref{fig:num_k8a002_t}, while $\omega \lesssim \omega_f$ in figure~\ref{fig:num_k0a002_t}. In summary, although the oscillating pattern is not as simple as that in the original coordinate space's Green's function due to the presence of $\omega$, we can find the quasinormal mode dependence in the spectral function. In appendix~\ref{app:time}, we compare the time dependence among different $k$ and give comments.

In the $\omega$-dependence at constant $\bar{t}$, first, the oscillation peaks align in a power law in intermediate $\omega$, and the alignment changes to exponential one in larger $\omega$. This point will be discussed in detail when we compare different $\Delta v$. Among these plots, the power law region looks wider in higher $k$. Second, as visible in the green and red lines in figures~\ref{fig:num_k0a002_w} and \ref{fig:num_k4a002_w}, the slope fluctuates in time, related to the phase shift in figures~\ref{fig:num_k0a002_t} and \ref{fig:num_k4a002_t}. The fluctuation is not clear in figure~\ref{fig:num_k8a002_w}, and in fact, such a phase shift is not apparent in figure~\ref{fig:num_k8a002_t}.

\begin{figure}[t]
\centering
\subfigure[$k=0$: $\omega=2$ (red), 10 (green), 40 (purple)]{\includegraphics[height=4.5cm]{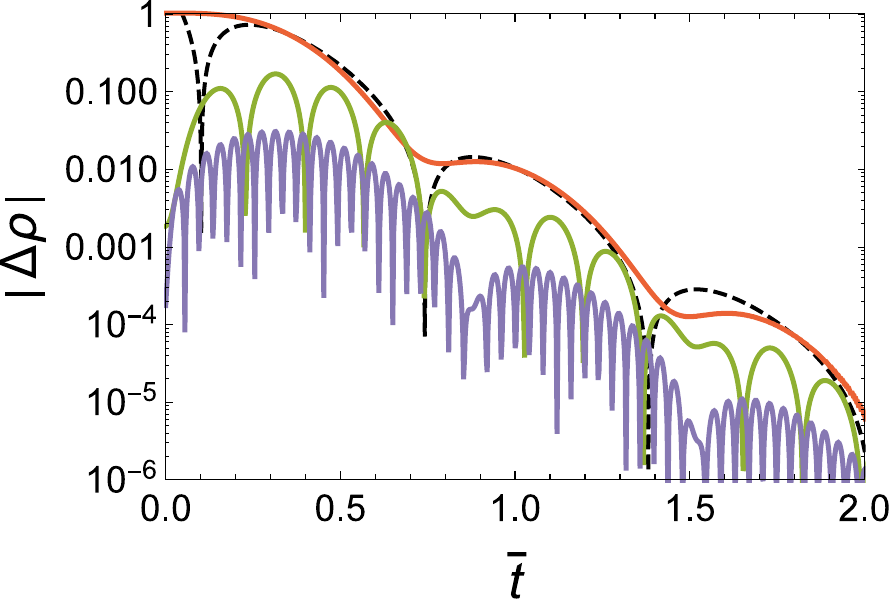}\label{fig:num_k0a001_t}}
\subfigure[$k=0$: $\bar{t}=0.4$ (blue), 0.8 (green), 0.9 (red)]{\includegraphics[height=4.5cm]{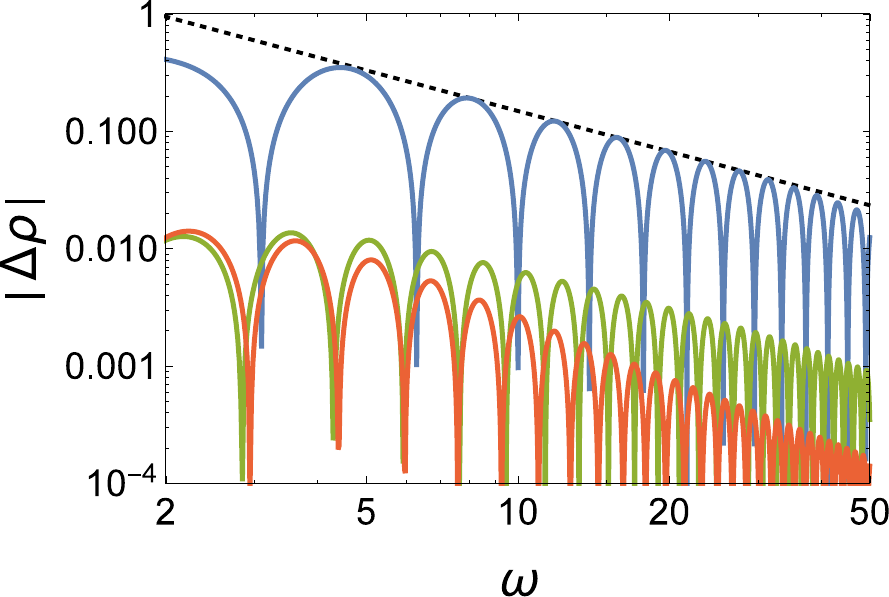}\label{fig:num_k0a001_w}}
\caption{Plots of $\Delta \rho$ when $\Delta v=0.01$. The black dashed line in the left panel is the same as that in figures~\ref{fig:num_k0a002_t}. The black dotted line in the right panel is $2.1 \, \omega^{-1.15}$.}
\label{fig:num_a001}
\end{figure}

Comparing the above results with different $\Delta v$, we argue that this timescale lies in the nonequilibrium spectral function. Firstly, plots for $\Delta v=0.01$ are shown for $k=0$ in figure~\ref{fig:num_a001}. We do not find a remarkable difference in the $\bar{t}$-dependence. In the $\omega$-dependence in figure~\ref{fig:num_k0a001_w}, the power law region extends to higher $\omega$ than the $\Delta=0.02$ case. It is plausible that the size of the power law range is related to $\Delta v$. In fact, the frequency $\omega$ can probe the width of the mass shell. If $\omega$ is too small, such a long wavelength does not probe the small $\Delta v$, and we would obtain the power law. If $\omega$ is large enough, such a high frequency probes the size of $\Delta v$ and the behavior would be exponential. The power seems to approach $\omega^{-1}$ as $\Delta v$ decreases. In the next section, we will reproduce this dependence.

\begin{figure}[t]
\centering
\subfigure[$\omega=5$ (red), 12 (green), 36 (purple)]{\includegraphics[height=4.5cm]{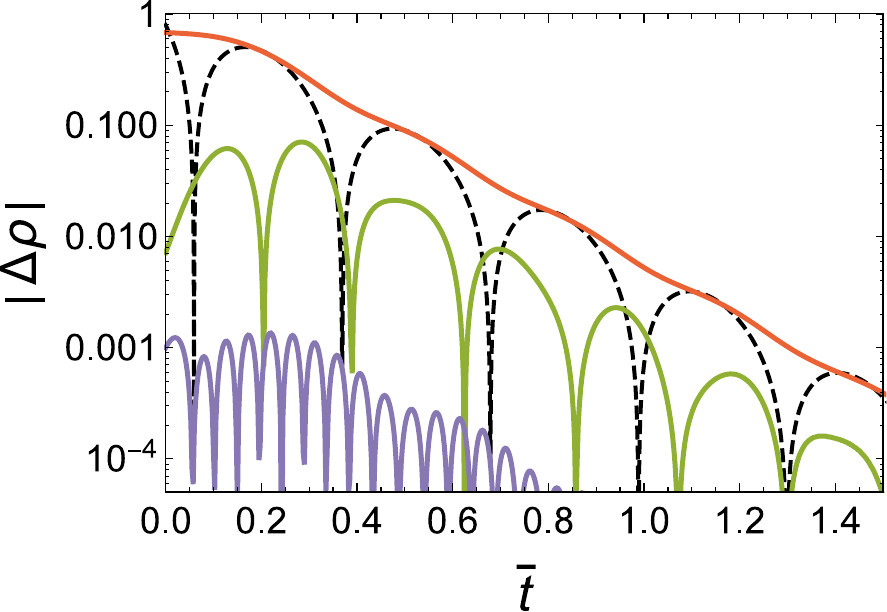}\label{fig:num_k4a005_t}}
\subfigure[$\bar{t}=0.4$ (blue), 1 (green), 1.1 (red)]{\includegraphics[height=4.5cm]{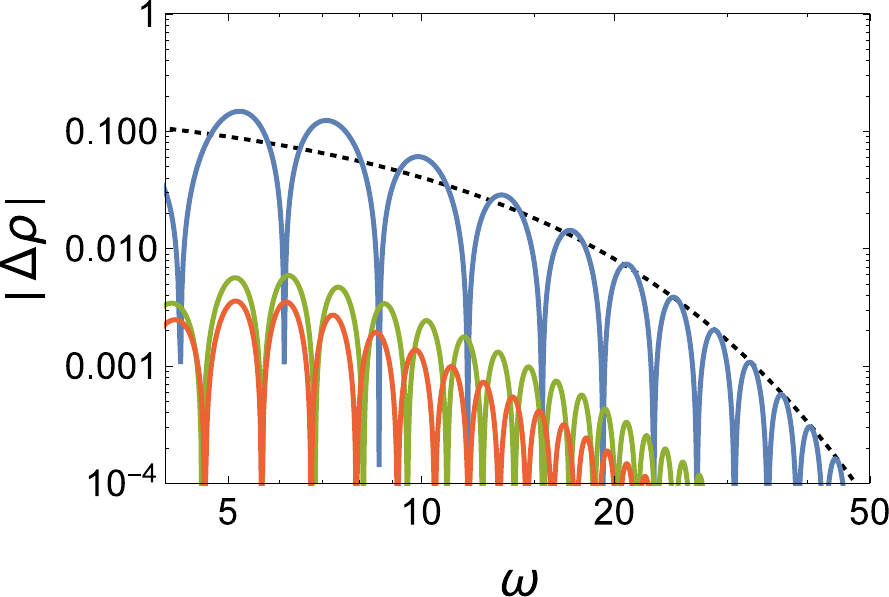}\label{fig:num_k4a005_w}}
\caption{Plots of $\Delta \rho$ for $k=4$ when $\Delta v=0.05$. The black dashed line in the left panel is the same as that in figure~\ref{fig:num_k4a002_t}. The black dotted line in the right panel is $0.2 \, e^{-0.16 \omega}$.}
\label{fig:num_a005}
\end{figure}

In figure~\ref{fig:num_a005}, we show results for a wider mass shell $\Delta v=0.05$ at $k=4$.
In the $\bar{t}$-dependence, the modulation with respect to $\omega_f$ is milder than that in the smaller $\Delta v$ case.
The shape of each peak also looks becoming featureless, maybe because of finite $\Delta v$ effects. In figure~\ref{fig:num_k4a005_w}, the power law $\omega$-dependence region is smaller and the exponential range is bigger compared with  figure~\ref{fig:num_k4a002_w}.

\section{A toy model motivated by holography}
\label{sec:toy}

In the previous section, we found several features that appear to depend on the frequencies and timescales in the system. To focus on their essential aspects, in this section, we consider a toy model motivated by the lowest quasinormal mode behavior. Some dynamics that may not be relevant is ignored. Nevertheless, main features in the previous section can be naively obtained.

The quasinormal mode behavior is nothing but damped oscillation. Therefore, we use an analogy to a damped harmonic oscillator driven by an impulse force,
\begin{equation}
m \ddot{x}(v) + 2 \gamma \dot{x}(v) + k x(v) = F_0 \delta(v-v_0),
\label{eq_fdho}
\end{equation}
where $\dot{x} \equiv \partial_v x$. For our use, we assume that the oscillator is underdamped: $\gamma^2 < k m$.\footnote{Generalization to overdamping is straightforward.} The response function (Green's function) is introduced as $x(v) = F_0 R(v,v_0)$. The solution is given by
\begin{equation}
R(v,v_0) = A_0 \sin (\omega_0 (v-v_0)) e^{-\Gamma_0 (v-v_0)} \theta(v-v_0),
\end{equation}
where $A_0=1/\omega_0$, $\omega_0=\pm \sqrt{km-\gamma^2}/m$ and $\Gamma_0 = \gamma/m$.

We use the above behavior to build a toy model for our temperature changing system. For simplicity, we firstly consider the case that the temperature changes instantaneously from $T_i$ to $T_f$ at $v=0$, corresponding to the thin-shell Vaidya-AdS, $M(v)=M_i+(M_f-M_i)\theta(v)$. We use the quasinormal mode value (table~\ref{QNMfreqtable}) for the frequency of the damped harmonic oscillator. To construct nonequilibrium oscillation, we connect the damped oscillators in the initial and final temperatures at $v=0$ smoothly.

This construction has some idealization compared with the actual time evolution done in the previous section. We ignore the initial nontrivial dynamics at the source application where higher quasinormal modes could contribute \cite{Heller:2013oxa,Gursoy:2016tgf}. Transition at $v=0$ is also trivialized, where it would actually take a finite time for the bulk field to rearrange. Nevertheless, these differences could be relevant only around $\bar{t} \simeq 0$ and will not be essential to the late time equilibrating pattern.

We thus consider a nonequilibrium generalization of the damped harmonic oscillator given as
\begin{equation}
\mathfrak{f}(v_2,v_1) =
\begin{cases}
A_i \theta(v_2-v_1) \Big( \theta(-v_2) \sin (\omega_i (v_2-v_1)) e^{- \Gamma_i (v_2-v_1)} & \\
\hspace*{\fill} + \alpha \, \theta(v_2) \sin (\omega_f v_2 + \delta) e^{- \Gamma_f v_2} \Big) & (v_1<0) \\
A_f \theta(v_2-v_1) \sin (\omega_f (v_2-v_1)) e^{- \Gamma_f (v_2-v_1)} & (v_1>0)
\end{cases}.
\label{toyGR_theta}
\end{equation}
Compared with the physical Green's function, the contribution from the delta function source at $v=v_1$ is not included because that contribution eventually cancels in the subtracted function $\Delta \rho$. The parameters $A, \, \omega$, and $\Gamma$ are the amplitude, the real part of the lowest quasinormal mode frequency, and that of the imaginary part, respectively, and the indices $i$ and $f$ indicate the initial and final temperature states. The factors $\alpha$ and $\delta$ specify the relative amplitude and phase across the thin shell. These are fixed by requiring the smoothness of $\mathfrak{f}$ and $\partial_v \mathfrak{f}$ at $v=0$, 
\begin{equation}
\alpha = -\frac{\sin (\omega_i v_1) e^{-\Gamma_i v_1}}{\sin \delta}, \quad
\delta = \cot^{-1} \left( \frac{\Gamma_f - \Gamma_i - \omega_i \cot(\omega_i v_1)}{\omega_f} \right).
\end{equation}
 
The parameters $A, \, \omega$, and $\Gamma$ are naturally related across $v=0$. We first assume the case corresponding to $k=0$. Finite $k$ will be commented later. The $k=0$ initial and final frequencies are related by $\omega_f = \omega_i(T_f/T_i)$ and $\Gamma_f = \Gamma_i(T_f/T_i)$. The relation for the amplitudes is $A_f = A_i(T_f/T_i)^2$ for the present $\Delta=2$ case. Generalization to other dimensions is straightforward.\footnote{Ultimately, the amplitude ratio will turn out unimportant for the features we will discuss, but for the time being we use this relation suggested by the numerical results.} The input for $A_i$ could be also fixed by comparing with the numerical results, but we may choose some $A_i$.
 
\begin{figure}[t]
\centering
\includegraphics[height=4cm]{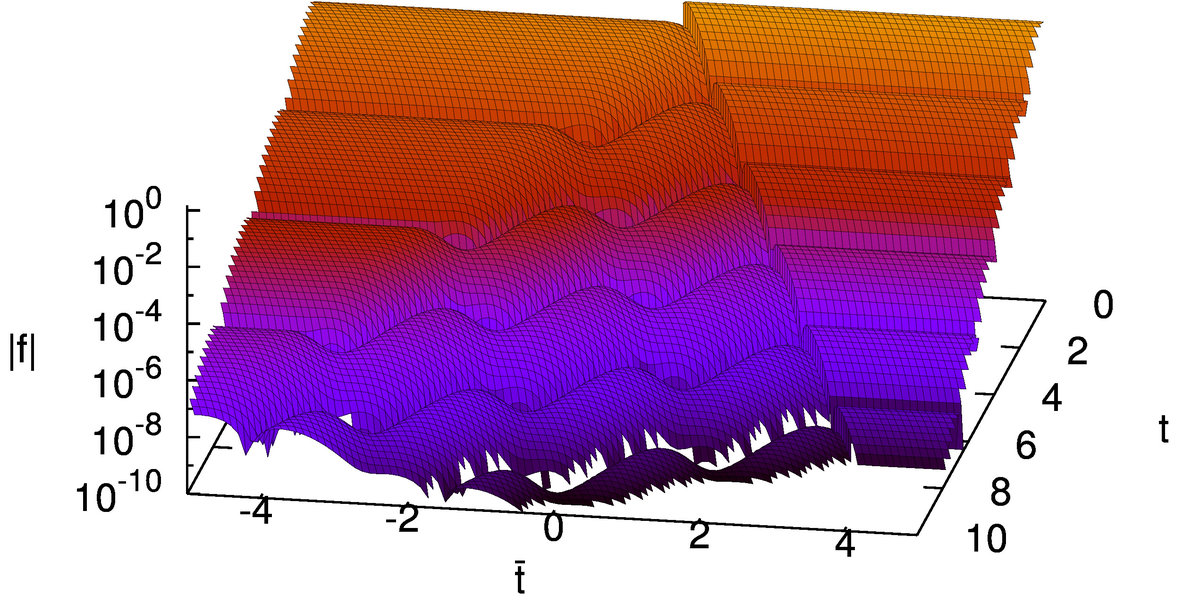}
\caption{Plot of $|\mathfrak{f}|$ in the $(t,\bar{t})$-coordinates. The thin-shell locates at $t=\pm 2 \bar{t}$.}
\label{fig:toyGR}
\end{figure}

The function \eqref{toyGR_theta} resembles well the Green's function. In figure~\ref{fig:toyGR}, $\mathfrak{f}(t,\bar{t})$ is plotted when $A_i=1$. This plot visualizes how the $\bar{t}$-dependence is in the $t$-integration in the Wigner transform. In particular, the equilibration comes from the change of $\mathfrak{f}$ at $t=2 \bar{t}$ where the background thin shell locates.\footnote{The other location of the mass shell is at $t=-2 \bar{t}$ that gives advanced changes to the spectral function since the Wigner transform is nonlocal in time.}

An benefit in \eqref{toyGR_theta} is that the Wigner transform is calculated analytically. It is defined by
\begin{equation}
\tilde{\mathfrak{f}}(\omega,\bar{t}) = \int dt \, e^{i\omega t} \mathfrak{f}(\bar{t}+t/2,\bar{t}-t/2).
\label{f_wig_def}
\end{equation}
The resulting expression is lengthy, and we show the explicit form in $\bar{t}>0$ only,
\begin{align}
\tilde{\mathfrak{f}}(\omega,\bar{t}) =& \frac{A_f \omega_f}{\omega_f^2 + (\Gamma_f-i \omega)^2} 
+ e^{-2 (\Gamma_f-i \omega) \bar{t}}
 \left( -A_f\frac{\omega_f \cos(2 \omega_f \bar{t}) + (\Gamma_f-i \omega) \sin(2 \omega_f \bar{t})}{\omega_f^2 + (\Gamma_f-i \omega)^2} \right. \nonumber \\
&\left.+ \frac{4iA_i \omega_i(\Gamma_f-i \omega)(\gamma\omega_f \cos(2 \omega_f \bar{t}) + (\gamma^2 + \omega_i^2 - \omega_f^2) \sin(2 \omega_f \bar{t}))}{\omega_f(\gamma^2+(\omega_i+\omega_f)^2)(\gamma^2+(\omega_i-\omega_f)^2)} \right),
\label{toyGR_wig}
\end{align}
where $\gamma \equiv \Gamma_i+\Gamma_f-2i\omega$. It is apparent that the damping of $\tilde{\mathfrak{f}}$ is given by $e^{-2 \Gamma_f \bar{t}}$, and the oscillations are a mix of frequencies $2\omega$ and $2\omega_f$. These features are consistent with the numerical results. To discuss the relaxation toward the equilibrium, we focus on
\begin{equation}
\hat{\mathfrak{f}}(\omega,\bar{t}) \equiv \tilde{\mathfrak{f}}(\omega,\bar{t}) - \tilde{\mathfrak{f}}(\omega,\infty).
\end{equation}
In particular, the imaginary part $\mathrm{Im} \, \hat{\mathfrak{f}}$ mimics the equilibration of the spectral function (up to normalization).

\begin{figure}[t]
\centering
\subfigure[$\mathrm{Im} \, \hat{\mathfrak{f}}$]{\includegraphics[height=4.5cm]{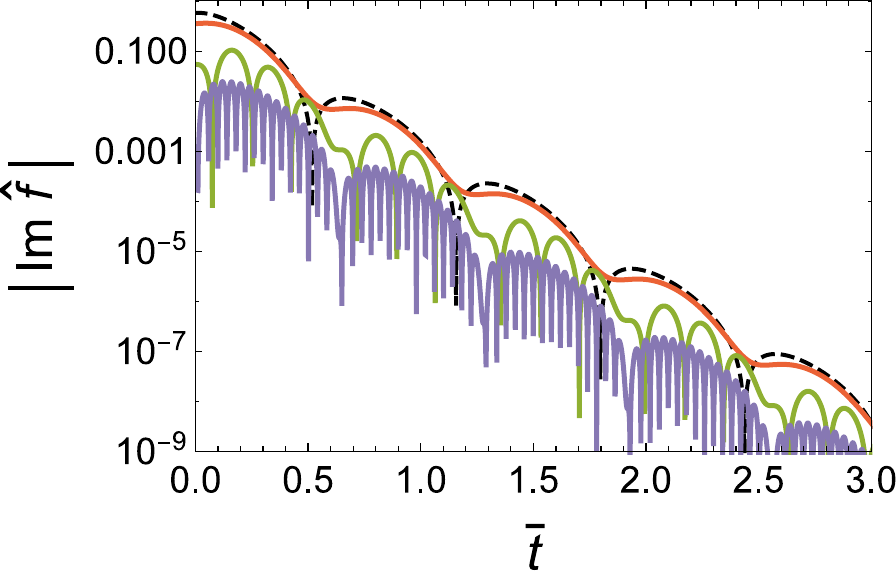}\label{f_im_t}}
\subfigure[$|\hat{\mathfrak{f}}|$]{\includegraphics[height=4.5cm]{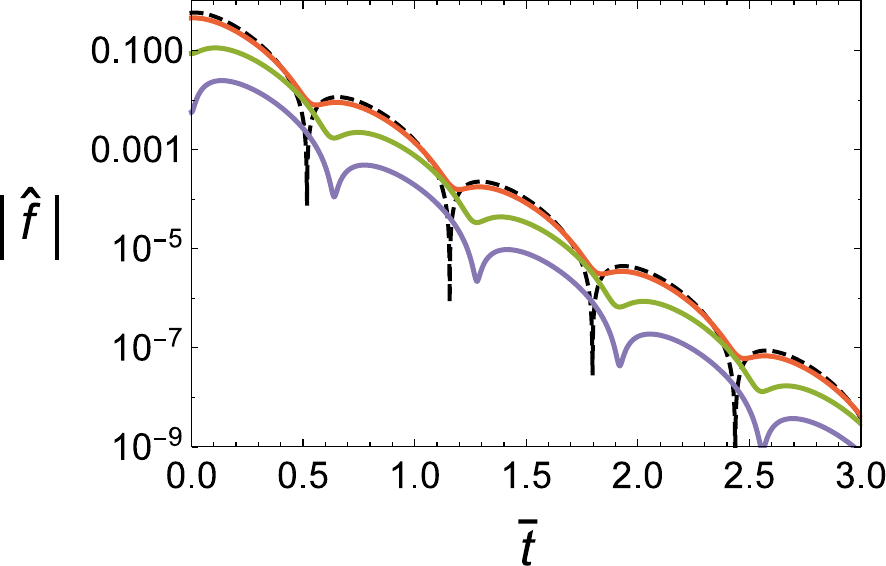}\label{fig:f_ab_t}}
\caption{The $\bar{t}$-dependence of $\mathrm{Im} \, \hat{\mathfrak{f}}(\omega,\bar{t})$ and $|\hat{\mathfrak{f}}|$ for $\omega/\omega_f=$ 1 (red), 4 (green), 16 (purple). Input parameters are $A_i=1$ and the $k=0$ quasinormal mode frequencies. The black dashed lines are both $\sin(2 \omega_f \bar{t}+0.6) e^{-2 \Gamma_f \bar{t}}$.}
\label{fig:fig_imf}
\end{figure}

The $\bar{t}$-dependence of $\mathrm{Im} \, \hat{\mathfrak{f}}$ is shown in figure~\ref{f_im_t}. In late $\bar{t}$, this is very similar to figure~\ref{fig:num_k0a001_t}. Difference in $\bar{t} \ll 1$ would be due to the absence of the (higher quasinormal mode) dynamics. The oscillations with respect to $\omega$ can be simply removed in $|\hat{\mathfrak{f}}|$, shown in figure~\ref{fig:f_ab_t}. This is because the $\omega$-oscillations from $e^{i \omega \bar{t}}$ cancel, and there remains an $\omega$-dependent phase shift.

\begin{figure}[t]
\centering
\includegraphics[height=4.5cm]{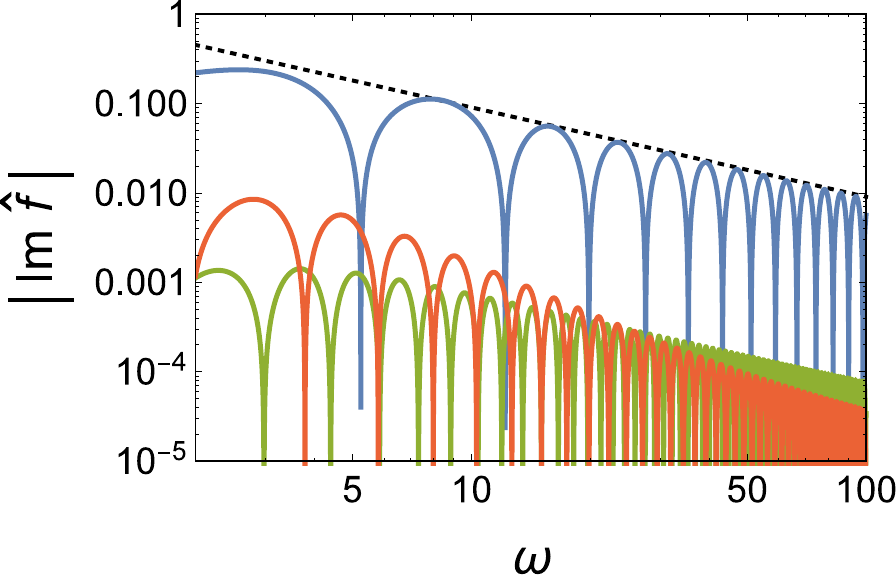}
\caption{The $\omega$-dependence of $\mathrm{Im} \, \hat{\mathfrak{f}}(\omega,\bar{t})$ at $\bar{t}=0.2$ (blue), 0.65 (green), and  1 (red). The black dotted line is $0.9 \, \omega^{-1}$.}
\label{fig:f_im_w}
\end{figure}

Main features in the $\omega$-dependence can be also reproduced. In figure~\ref{fig:f_im_w}, we show the $\omega$-dependence of $\mathrm{Im} \, \hat{\mathfrak{f}}$. The peaks align with the power law given by $\omega^{-1}$. The slope fluctuates corresponding to the phase shift in figure~\ref{fig:f_ab_t}. The fluctuation repeats each $\bar{t}_\mathrm{period} = \pi/(2\omega_f) =0.64$. Difference from the $\Delta v=0.01$ case in figure~\ref{fig:num_k0a001_w} would be mainly due to finite $\Delta v$ effects.

This toy description can be applied to finite $k$. We may use numerical values for the initial and final quasinormal mode frequencies. The $k$-dependence in the amplitude is nontrivial. But, it mainly affects the normalization and is not essential in the equilibrating pattern. We could simply impose $A_i=A_f$ as if the momentum overwhelms temperature.

We can even consider a finite $\Delta v$ generalization. In terms of the frequencies, \eqref{eq_fdho} is written as
\begin{equation}
\ddot{R} + 2 \Gamma_0 \dot{R} + \kappa R = \alpha \delta(v-v_0),
\label{eq_fdho_alt}
\end{equation}
where $\dot{R} \equiv \partial_v R$, $\kappa=\omega_0^2 + \Gamma_0^2$ and $\alpha = 1/m$. We generalize this equation to time dependent coefficients. Assuming $k=0$, we scale the dimensionful quantities with the background temperature. We also make $R$ and $\alpha$ in \eqref{eq_fdho_alt} to have a different mass dimension ($\Delta=2$ here). Thus, we consider
\begin{equation}
\ddot{\mathfrak{g}}(v) + 2 \Gamma_0(v) \dot{\mathfrak{g}}(v) + \kappa(v) \mathfrak{g}(v) = \alpha(v) \delta(v-v_0),
\label{eq_noneq_fdho}
\end{equation}
where $\dot{\mathfrak{g}} \equiv \partial_v \mathfrak{g}$, and the time dependent coefficients are
\begin{equation}
\Gamma_0(v) = \Gamma_i T(v), \quad \kappa(v) =  (\omega_i^2 + \Gamma_i^2)T(v)^2, \quad \alpha(v) = \alpha_i T(v)^{\Delta+1},
\label{param_noneq_fdho}
\end{equation}
where we use $T(v) = (M(v)/M_i)^{1/3}$ giving the instantaneous temperature ratio as if the background change is quasistatic. We use the $k=0$ quasinormal mode frequency for $\omega_i$ and $\Gamma_i$. For the amplitude, we try $\alpha_i=1$.

We solve \eqref{eq_noneq_fdho} numerically because we did not find an analytic solution. We focus on the equilibration
\begin{equation}
\hat{\mathfrak{g}}(\omega,\bar{t}) \equiv \tilde{\mathfrak{g}}(\omega,\bar{t}) - \tilde{\mathfrak{g}}(\omega,\infty).
\end{equation}
where $\tilde{\mathfrak{g}}(\omega,\bar{t})$ is the Wigner transform of $\mathfrak{g}$ defined in the same way as \eqref{f_wig_def}.\footnote{Actually, $\mathfrak{g}$ is the two point function having the arguments as $\mathfrak{g}(v,v_0)$.}

\begin{figure}[t]
\centering
\includegraphics[height=4.5cm]{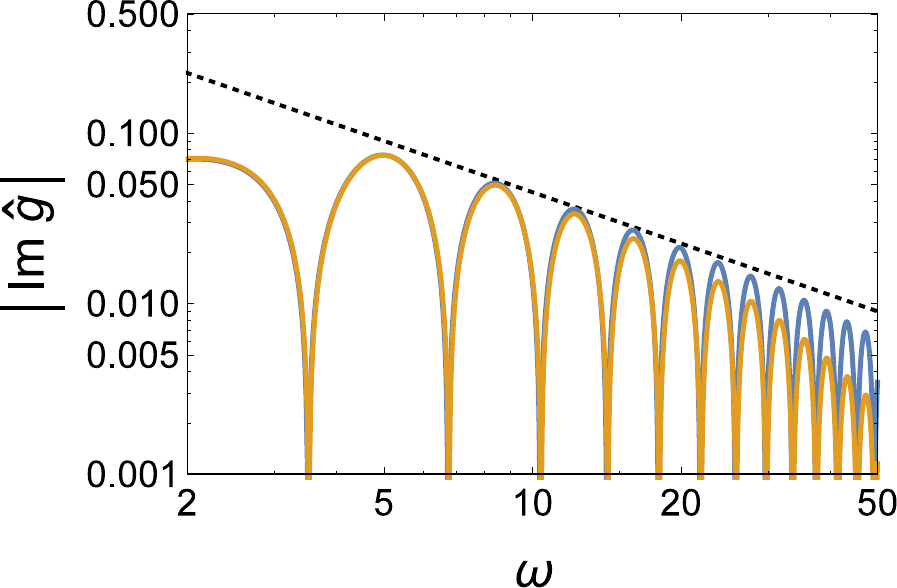}
\caption{The $\omega$-dependence of $\mathrm{Im} \, \hat{\mathfrak{g}}$ for $\Delta v=$0.01 (blue), 0.02 (orange) when $\bar{t}=0.4$. The black dashed line is $0.45 \, \omega^{-1}$.}
\label{fig:g_compare}
\end{figure}

In figure~\ref{fig:g_compare}, the $\omega$-dependence for $\Delta t=0.01$ and 0.02 are compared. The power-law/exponential feature is found. In the $\Delta \to 0$ limit, $\mathfrak{g}$ approaches the $\omega^{-1}$ power law. In fact, the thin-shell model is obtained in that limit. The $\Delta v$ dependence hence is simply described by the kind of model given in here. It seems straightforward to generalize this to finite $k$, but we need to specify nontrivial $k$-dependence.

Thus, in the toy model by the generalized forced damped harmonic oscillator, we can find main features in the equilibration of the spectral function. This might imply that the equilibration of the (frequency-analyzed) correlation function is rather simple in the sense that it is fairly well dominated by the lowest quasinormal mode behavior, and other dynamics that is ignored here but exists in the full evolution would not alter the behavior drastically. The full time evolution would be of more interest in early time far-from-equilibrium dynamics. There may be minor differences that are not included in the toy model, but these are not of main interest of the aspects we focused on.

\section{Summary}
\label{sec:dis}

We studied the ringdown of the nonequilibrium Green's function in a theory with the holographic dual in detail, as a complement to the preceding work \cite{Banerjee:2016ray}. We provided further numerical computations in Vaidya-AdS geometry with the narrower mass shell than the preceding work. It was confirmed that the nonequilibrium spectral function decayed with the final state's lowest quasinormal mode frequency. The damping rate was set by the frequency's imaginary part, and the time oscillations depended on the real part of the frequency and the Wigner frequency $\omega$. We also showed that the timescale $\Delta v$ is encoded in the spectral function. The $\omega$-dependence approached the power law as $\Delta v$ is decreased. In appendix~\ref{app:time}, $k$-dependence in the equilibrating spectral function were discussed.

We also considered a toy model motivated by the quasinormal mode behavior and the forced damped harmonic oscillator. We considered a nonequilibrium generalization of the oscillator to the temperature changing system. While this model ignored some dynamics in the actual physical time evolution and rather traced the quasistatic frequency change, main features in the equilibrating spectral function were simply realized. To conclude, the lesson is that the quasinormal mode behavior is important in (holographic) correlation functions, as expected.

Because our toy model relies only on the presence of the damping oscillation due to the lowest quasinormal mode frequency, it is expected that our description would be generally useful in other operators. In this paper, we had mainly in mind a scalar operator. It would be interesting to discuss the behaviors of different operators from our simple viewpoint. We speculate essential aspects in the equilibration of strongly coupled systems are fairly simple such that the lowest quasinormal mode dependence is dominant.

\acknowledgments
The author thanks Souvik Banerjee, Lata Kh Joshi, Ayan Mukhopadhyay, and P. Ramadevi for collaboration in the early stage of this work and comments on the manuscript. The author also thanks Matthias Kaminski, Johannes Oberreuter, Paul Romatschke, Christopher Rosen, and Jackson Wu for useful discussions. This work was supported by the Department of Energy, DOE award No.~DE-SC0008132.

\appendix

\section{Momentum dependence in the spectral function's equilibration}
\label{app:time}
\begin{figure}[t]
\centering
\includegraphics[height=4.5cm]{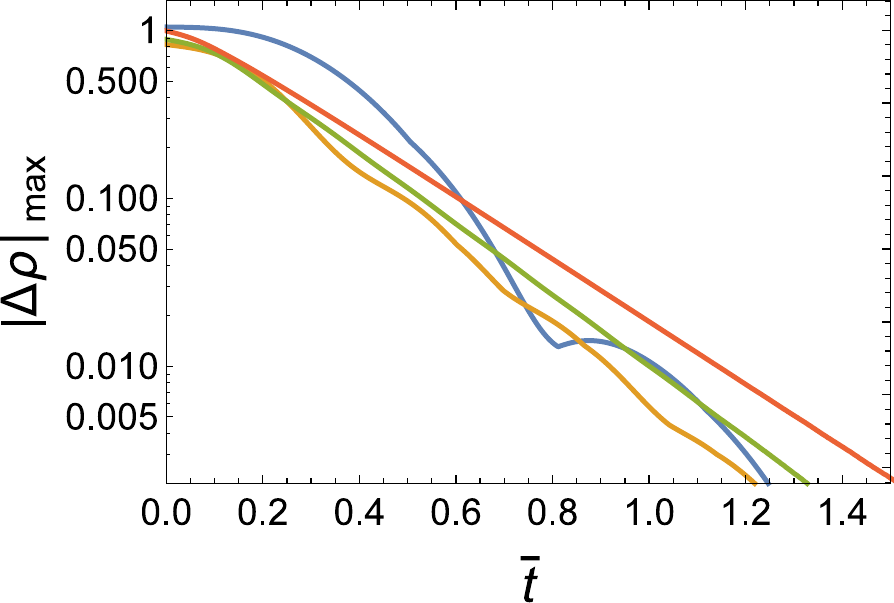}
\caption{$|\Delta \rho(\bar{t};\bm{k})|_\mathrm{max}$ for $\Delta v=0.02$ with $k=0$ (blue), 4 (orange), 8 (green), 16 (red).}
\label{fig:rhomax}
\end{figure}

In this appendix, we discuss how the spectral function is different from the final state in different $k$. As a measure, we consider the maximal difference of the spectral function from the final configuration at each $\bar{t}$,\footnote{One may use a different measure and obtain different conclusions than this appendix.}
\begin{equation}
|\Delta \rho(\bar{t};\bm{k})|_\mathrm{max} \equiv \max_{\omega>0} \left( |\Delta \rho(\omega,\bar{t};\bm{k})|\right).
\label{rhomaxdiff}
\end{equation}
Evaluating this, we find that the frequencies that mainly contribute are around $\omega \simeq \omega_f$. Thus, $|\Delta \rho|_\mathrm{max}$ is tied to the amount of excited lowest quasinormal mode. In figure~\ref{fig:rhomax}, we compare $k=0$, 4, 8, 16 for $\Delta v =0.02$.\footnote{We use the same $T_i=3/(4\pi)$ and $T_f/T_i=2$ as the main text.} The decay in each plot is in fact given by the imaginary part of each final state's lowest quasinormal mode frequency.\footnote{The final state quasinormal mode frequency for the $k=16$ case is $\omega_f - i\Gamma_f = 16.73 - 2.147i$.} The initial short time before each plot starts to decrease would be considered as the timescale for nontrivial early time dynamics other than the quasinormal mode equilibration.

In the plots, the magnitude in small $\bar{t}$ is the biggest at $k=0$ and smaller in finite $k$. However, the change is not monotonic in $k$. It initially decreases as $k$ is turned on, becomes the minimum around $k \simeq 3$, and moderately increases in $k \gtrsim 3$.\footnote{If $\Delta v$ is varied a little, results hardly change. We did not try varying $T_f/T_i$. Therefore, results could be quantitatively different from the present ones (plotted at $k/(4\pi T_f/3)=0,\,2,\,4, \, 8$) if the temperature ratio is changed.} As a result, the plots in $k\ge 4$ are not suppressed around $\bar{t} \simeq 0$. In small $k$, the decrease of the initial magnitude dominantly contributes to the difference in finite $\bar{t}$ as we can find between the $k=0$ and $4$ plots. However, once the decrease stops, the quasinormal mode dependence is more relevant giving a slower decay in higher $k$ as we see in the $k=4, \, 8, \, 16$ plots.\footnote{This observation is when \eqref{rhomaxdiff} is considered and could be different if a different measure is used. Also, if the momentum dependence is absent such as in $AdS_3$, a different outcome would apply.}

We may define a thermalization time at which the difference of the spectral function from the final state is smaller than a cutoff. In the literature, such a cutoff time is often used to argue when the nonequilibrium evolution would be practically close to the thermal equilibrium (e.g.~\cite{Chesler:2008hg,Heller:2012km,Buchel:2012gw}). In figure~\ref{fig:rhomax}, such a time may be determined by drawing a horizontal line and seeing the intersection of the decaying line with it. In \cite{Banerjee:2016ray}, a shorter thermalization time for the spectral function was claimed when a (small) finite $k$ was turned on.\footnote{The study in \cite{Banerjee:2016ray} was limited to $k \le 4.5$. Generally on thermalization in large momentum, see also \cite{Balasubramanian:2011ur} for early comments and \cite{Keranen:2015mqc} for further discussion.} In broad strike, in figure~\ref{fig:rhomax}, this tendency can be seen in comparing $k=0$ and $k=4$ plots that have different initial magnitudes but the decay slopes are not so much different.\footnote{A precise comparison including $k\simeq 3$ and using a small $k$ step could give involved results around that momentum as notified in the previous paragraph.}

Once the momentum dependence does not suppress the overall magnitude, however, the quasinormal mode timescale can be more relevant. This is seen in the $k=4, \, 8, \, 16$ plots in figure~\ref{fig:rhomax}. The quasinormal mode dominance is more apparent in higher $k$ where the imaginary part of the frequency is closer to zero.

\bibliography{bib_toyGR}

\end{document}